\documentclass[12pt]{article}
\usepackage{graphicx}
\usepackage{amsfonts,amsthm}
\usepackage{amsmath,amssymb}
\usepackage{bm}
\usepackage[utf8]{inputenc}
\usepackage[usenames,dvipsnames]{xcolor}

\newcommand{\be}{\begin{equation}}
\newcommand{\ee}{\end{equation}}
\newcommand{\bea}{\begin{eqnarray}}
\newcommand{\eea}{\end{eqnarray}}
\newcommand{\dst}{\displaystyle}
\newcommand{\lr}[1]{ \langle #1 \rangle}

\newcommand{\fr}[2]{\frac{{\dst #1}}{{\dst #2}}}

\newcommand{\mmmatrix}[9]{ \left(\! \begin{array}{ccc}#1 & #2 & #3\\[1mm] #4 & #5 & #6\\[1mm] #7 & #8 & #9\\ \end{array}\!\right) }

\newcommand{\triplet}[3]{ \left( \begin{array}{c}#1 \\[2mm] #2 \\[2mm] #3\end{array}\right) }

\newcommand{\bA}{{\bf A}}

\newcommand{\bbe}{{\bf e}}
\newcommand{\bk}{{\bf k}}
\newcommand{\bn}{{\bf n}}
\newcommand{\bp}{{\bf p}}

\newcommand{\br}{{\bf r}}

\def\lsim{\mathrel{\rlap{\lower4pt\hbox{\hskip1pt$\sim$}}
    \raise1pt\hbox{$<$}}}         
\def\gsim{\mathrel{\rlap{\lower4pt\hbox{\hskip1pt$\sim$}}
    \raise1pt\hbox{$>$}}}         

\title{
{\normalsize \hfill CFTP/19-015} \\*[7mm]
Fate of the Landau-Yang theorem for twisted photons}

\author{Igor~P.~Ivanov$^1$, Valeriy G. Serbo$^{2,3}$, Pengming Zhang$^{4,5}$
\\
{\small $^1$ CFTP, Instituto Superior Tecnico, Universidade de Lisboa, Lisbon 1049-001, Portugal}\\
{\small $^2$ Novosibirsk State University, 630090, Novosibirsk, Russia}\\
{\small $^3$ Sobolev Institute of Mathematics, 630090, Novosibirsk, Russia}\\
{\small $^4$ School of Physics and Astronomy, Sun Yat-sen University, Zhuhai 519082, China}\\
{\small $^5$ Institute of Modern Physics, Chinese Academy of Sciences, Lanzhou 730000, China}
}

\begin{document}

\maketitle

\begin{abstract}
Landau-Yang theorem is sometimes formulated as a selection rule
forbidding two real (that is, non-virtual) photons with zero total momentum to be in the state of the total angular momentum $J=1$.
In this paper we discuss whether the theorem itself and this particular formulation can be extended
to a pair of two {\em twisted} photons, which carry orbital angular momentum
with respect to their propagation direction.
We point out possible sources of confusion, which may arise both
from the unusual features of twisted photons and from the fact
that usual proofs of the Landau-Yang theorem operate in the center of motion reference frame,
which, strictly speaking, exists only for plane waves.
We show with an explicit calculation that a pair of twisted photons
does have a non-zero overlap with the $J=1$ state.
What is actually forbidden is production of a spin-1 particle by such a photon pair,
and in this formulation the Landau-Yang theorem is rock-solid.
Although both the twisted photon pair and the spin-1 particle can exist in the $J=1$ state,
these two systems just cannot be coupled in a gauge-invariant and Lorentz invariant manner respecting Bose symmetry.
\end{abstract}

\section{The problem}

Twisted, or vortex, photons are monochromatic, freely propagating solutions of Maxwell's equations
with helicoidal phase fronts, see classical works \cite{Allen:1992zz,Allen:1999}
and recent reviews \cite{Molina-Terriza:2007,Zhan:2009,Paggett:2017,Knyazev:2019}.
Colloquially speaking, such photons rotate around the propagation direction
and carry non-zero intrinsic orbital angular momentum (OAM).
This is not a collective effect; after quantization, each photon carries a non-zero OAM.

Since twisted photons carry OAM, one may wonder whether a pair of such photons could bypass
the restrictions imposed by the famous Landau-Yang theorem \cite{Landau:1948kw,Yang:1950rg}.
This theorem has immediate consequences in particle physics,
such as, for example, the strongly enhanced lifetime of orthopositronium due to the absence
of the two-photon decay channel.
Thus, if there could exist any loophole in the Landau-Yang theorem, it will have dramatic effect.

One can encounter in literature two formulations of the Landau-Yang theorem:
\begin{enumerate}
\item
a spin-1 particle cannot decay into nor be produced by a pair of real photons;
\item
two real plane wave photons with zero total momentum
cannot be found in a state with the total angular momentum $J=1$.
\end{enumerate}
The two formulations are considered completely equivalent, and the actual proofs often switch from one formulation
to the other, making use of the plane wave configurations in the second formulation.

It seems that, at least, the {\em language} of both formulations is well suited to twisted photon collisions.
The questions we want to address are whether the Landau-Yang theorem indeed applies to twisted photons
and whether the two formulations need (and can) be adapted to this case.
It is already here, at first thinking, that certain confusion may arise.

On the one hand, if the Landau-Yang theorem forbids certain production or decay process
for any plane wave configuration (formulation 1), then one cannot overcome this restriction by any engineering
of the initial or final wave packets. Indeed, the amplitude of the orthopositronium decay in an arbitrary spatial wave function
to a pair of twisted photons can always be expanded in terms of plane wave amplitudes and must eventually produce zero.
This result seems to be rock-solid.

On the other hand, it is not immediately obvious whether the angular momentum selection rule
in formulation 2 can be extended to twisted photons.
There exist several ways to prove the Landau-Yang theorem
\cite{Landau:1948kw,Yang:1950rg,LL4,Beenakker:2015mra,Cacciari:2015ela,Pleitez:2015cpa}.
But all of them make use of simplifications which explicitly rely on plane waves
and which do not hold for twisted photons. To make this point clearer,
we give a list of considerations which may stir doubt.
\begin{itemize}
\item
Since a twisted photon is not a plane wave, it is not characterized by a well-defined
four momentum $k^\mu$ nor by a well-defined, coordinate-independent polarization vector $e^\mu$.
One can introduce the {\em average} 3D momentum $\lr{\bk}$ but it does not satisfy the usual dispersion law
$E^2 - \lr{\bk}^2 \not = 0$, which is tempting to interpret as an ``effective mass'' of the twisted photon.
Since we know that two {\em massive} spin-1 particles, including the case of virtual photons,
can bypass the Landau-Yang theorem, it may make one wonder if a similar phenomenon could happen to twisted photons.
\item
When proving the Landau-Yang theorem, one usually switches to the center of motion reference
frame, in which the two photons are back to back:
\be
k_1^\mu = E (1, 0, 0, 1)\,, \quad k_2^\mu = E (1, 0, 0, -1)\,. \label{cmf}
\ee
In this frame, any massive particle which emerges from the $\gamma\gamma$ fusion will be at rest.
But for twisted photons, this does not hold. A twisted photon is a steady interference pattern of various plane wave
components with different (transverse) momenta.
Even if one selects a reference frame where the sum of the average momenta is zero,
\be
\lr{\bk_1} + \lr{\bk_2} = 0\,,\label{balanced}
\ee
a particle emerging in fusion of two twisted photons will still display
a distribution over a certain range of final momenta centered around zero.
It will never be at rest. Thus, the starting assumption in most proofs is not applicable anymore.
\item
For a pair of plane wave photons, when working in the c.m. frame \eqref{cmf}, one can make use of gauge freedom
to set $e_1^\mu$ orthogonal not only to $k_1^\mu$ but also to $k_2^\mu$.
As a result, when coupling the two-photon system to the final particle,
one can immediately neglect terms involving $(e_1 k_2) \equiv e_1^\mu k_{2\mu}$ and $(e_2 k_1)$.
For twisted photons, this does not automatically hold.
When two twisted photons collide, each plane wave component of the first photon with momentum $k_1$
sees a coherent superposition of the plane wave components of the second photon, each with its own $k_2$.
Thus, it is impossible to adjust the polarization vector $e_1$ of this plane-wave component
which would be simultaneously orthogonal to all plane-wave components of the second photon.
Therefore, terms involving $(e_1 k_2)$ need to be treated with care.
\item
One should not forget that twisted photons, being cylindrical beams, carry non-zero angular momentum
{\em projection} $J_z$. They are not eigenstates of $J$. One needs to check how it affects formulation 2.
\end{itemize}
In short, the real question is whether formulation 2 must (or can!) be adapted to twisted photons,
or whether it must be abandoned altogether.

The purpose of this Letter is to dissipate this confusion and, with direct calculations, provide the answers.
We will show that formulation 1 remains unchanged, while formulation 2 must be abandoned
because two twisted photons do have overlap with a state $J=1$,
even in the frame where the sum of their average momenta is zero.
We believe that our work will provide additional clarification and a novel pedagogical insight
into this rather old problem.

The structure of this paper is as follows. In the next section we recapitulate the description
of twisted photons and massive vector particles. In section~\ref{section-collision} we consider the process of
collision of two twisted photons and compute, first, production of a scalar particle and then of a spin-1 particle.
Finally, we draw our conclusions.
Throughout the paper, 3D vectors are given in bold such as $\bk$, 2D vectors carry $\perp$ subscript, such as $\bk_\perp$,
while the products of four-vectors are written as $(k_1 k_2) \equiv k_{1}^\mu k_{2\mu}$.

\section{Describing twisted vector fields}

In this section, we recapitulate the formalism of constructing vortex vector fields
and describing its polarization state. We do it first for the photons and then for massive vector particles.
This exposition is based on \cite{Jentschura:2010ap,Jentschura:2011ih} and later publications.

We begin with a monochromatic plane-wave electromagnetic field with helicity
$\lambda = \pm 1$, which is described, in the Coulomb gauge, by
\be
\bA_{\bk \lambda}(\br) = \bbe_{\bk \lambda}\, e^{i\bk \br}\,.\label{PW1}
\ee
The polarization vector is orthogonal to the wave vector: $\bbe_{\bk\lambda} \bk = 0$.
Quantization of this field produces plane wave photons with momentum $\hbar \bk$.
From now on, we switch to the natural units $\hbar = c = 1$.

Let us now fix a reference frame and select an axis $z$.
The simplest form of twisted photon a cylindrical Bessel photon with helicity $\lambda = \pm 1$
and total angular momentum $J_z = m$ propagating on average along axis $z$.
It is a monochromatic solution of Maxwell's equations which is constructed
as a superposition of plane waves with fixed longitudinal momentum
$k_z = |\bk|\cos\theta$, fixed modulus of the transverse momentum $\varkappa = |\bk_\perp| = |\bk|\sin\theta$,
but arriving from different azimuthal angles $\varphi_k$.
The usual dispersion relation holds for every plane wave component: $k_z^2 + \varkappa^2 = E^2$,
where $E$ is the energy of the photon.
Using the Coulomb gauge for all plane wave components, we get
\be
\bA_{\varkappa m \lambda}(\br) = \int a_{\varkappa m}(\bk_\perp)\, \bbe_{\bk \lambda}\, e^{i\bk \br} {d^2k_\perp \over (2\pi)^2}\,,
\label{tw1}
\ee
where the Fourier amplitude $a_{\varkappa m}(\bk_\perp)$ is given by
\be
a_{\varkappa m}(\bk_\perp) = i^{-m} e^{im\varphi_k} {2\pi \over \varkappa}\delta(k_\perp - \varkappa)\,.\label{a}
\ee
The Fourier amplitude is an eigenfunction of the $z$-projection of the orbital angular momentum operator
$\hat{L}_z = -i \partial/\partial \varphi_k$ with the eigenvalue $m$.

The polarization vector $\bbe_{\bk \lambda}$ inside the integral \eqref{tw1}
depends on $\bk$ and cannot be taken out of the integral. It means that the polarization state
of a twisted photon, as a whole, is described by a polarization {\em field} rather than polarization vector.
For each plane wave component of a twisted photon, the polarization vector is {\em not} an eigenstate of $\hat{L}_z$.
This is the origin of the spin-angular interaction inside a free photon, which results in
non-conservation of the OAM and spin projections separately \cite{Bliokh:2015yhi,Knyazev:2019}.
We add, however,
that in most experimental situations, the twisted photons are produced in the paraxial regimes,
where $\theta \ll 1$, and one can talk about approximately conserved $L_z$ and $s_z$ \cite{Allen:1992zz,Allen:1999}.

To describe the polarization vectors of photons with arbitrary momentum,
let us define the eigenvectors ${\bm \chi}_\sigma$, $\sigma= \pm 1, 0$,
of the spin $z$-projection operator $\hat{s}_z$: $\hat{s}_z {\bm \chi}_\sigma = \sigma {\bm \chi}_\sigma$.
There explicit form is
\be
   {\bm \chi}_{0}=
   \left(
   \begin{tabular}{c}
                 0 \\
                 0 \\
                 1 \\
   \end{tabular}
   \right),\
   {\bm \chi}_{\pm 1}= \fr{\mp 1}{\sqrt{2}}
   \left(
   \begin{tabular}{c}
                 1 \\
                 $\pm i$ \\
                 0 \\
   \end{tabular}
   \right)\,, \quad {\bm \chi}^*_\sigma {\bm \chi}_{\sigma^\prime} = \delta_{\sigma\sigma^\prime}\,.
   \label{chi}
\ee
The polarization vector can be expanded in the basis of ${\bm \chi}_\sigma$:
 \be
 \bbe_{\bk \lambda}=
 \sum_{\sigma=0,\pm 1} e^{-i\sigma \varphi_k}\,
 d^{1}_{\sigma \lambda}(\theta)  \,\bm \chi_{\sigma}\,.
\label{bbe-chi}
 \ee
The explicit expressions for Wigner's $d$-functions, which compactly describe a pure polarization state
in an arbitrary basis~\cite{LL3,Varshalovich}, are:
\be
d^{1}_{\sigma\lambda}(\theta) = \mmmatrix{\cos^2\fr{\theta}{2}}{-\fr{1}{\sqrt{2}}\sin\theta}{\sin^2\fr{\theta}{2}}%
{\fr{1}{\sqrt{2}}\sin\theta}{\cos\theta}{-\fr{1}{\sqrt{2}}\sin\theta}%
{\sin^2\fr{\theta}{2}}{\fr{1}{\sqrt{2}}\sin\theta}{\cos^2\fr{\theta}{2}}%
\ee
The first, second, and third rows and columns of this matrix correspond to the indices $+1,\, 0,\, -1$.
Performing the summation in Eq.~\eqref{bbe-chi}, one gets explicit expressions for the polarization vectors:
\be
\bbe_{\bk \lambda}= \fr{\lambda }{\sqrt{2}}\triplet{-\cos\theta\cos\varphi_k + i \lambda \sin\varphi_k}%
{-\cos\theta\sin\varphi_k - i \lambda \cos\varphi_k}{\sin\theta}\,,\quad \lambda= \pm 1\,.\label{bbe-explicit}
\ee
Notice that the vectors $\bbe_{\bk \lambda}$ are constructed in such a way that they
are eigenvectors of the $z$-component of the total angular momentum operator $\hat{J}_z = \hat{L}_z + \hat{s}_z$
with zero eigenvalue.
In the paraxial approximation, when $\theta\to 0$, this polarization vector becomes
\be
\bbe_{\bk \lambda} \to e^{-i\lambda \varphi_k} \,\bm \chi_{\lambda}\,,
\ee
which is still an eigenstate of $\hat{J}_z$ with zero eigenvalue.
If needed, one can explicitly perform the angular integration in Eq.~\eqref{tw1} and obtain
a compact expression in cylindrical coordinates
$\br = (\rho \cos\varphi_r, \rho \sin\varphi_r, z)$:
\be
\bA_{\varkappa m \lambda}(\br) = e^{ik_z z}  \sum_{\sigma = \pm 1, 0} i^{-\sigma} \, d^1_{\sigma \lambda}(\theta)\,
J_{m-\sigma}(\varkappa \rho) \, e^{i(m-\sigma)\varphi_r} {\bm \chi}_\sigma\,.
\ee
A counter-propagating twisted photon, defined in the same reference frame
with respect to the same axis $z$, can be described by the above expressions assuming that $k_z < 0$
and replacing $m \to - m$ in the Fourier amplitude \eqref{a}.
The expression for the polarization vector \eqref{bbe-explicit} stays unchanged,
but $\cos\theta < 0$. The paraxial limit is now given by $\theta \to \pi$, in which case
\be
\bbe_{\bk \lambda} \to e^{+i\lambda \varphi_k} \,\bm \chi_{-\lambda}\,,
\ee

Let us now discuss the gauge transformation freedom which exists for twisted photons.
In a generic gauge, one works with the four-potential $A_\mu$ rather than its space-like
part in Eqs.~\eqref{PW1} and \eqref{tw1}.
For a plane wave photon, one can always perform the shift
\be
e^\mu \to e^{\prime \mu} = e^\mu + c k^\mu\,,
\ee
where $c$ can depend on photon's momentum $k^\mu$ as well as on other momenta in the problem.
For a twisted photon, one is allowed to perform this gauge transformation under the integral \eqref{tw1},
independently for each plane wave component. In particular, if one wishes make $e^\mu$ orthogonal
to a different four-vector $p^\mu$ not parallel to $k^\mu$, then one can perform the following
gauge transformation
\be
e^\mu \to e^{\prime \mu} = e^\mu  - k^\mu {(e p) \over (k p)} \,,\label{gauge-tr}
\ee
so that $(e' p) = 0$. We stress again that this change takes place under the integral.


The above formalism can be immediately extended to a massive spin-1 particle of mass $M$
described with the polarization vector $V_\mu(\lambda_V)$.
The only modifications are that the dispersion relation changes
to $p_z^2+\varkappa^2+M^2 = E^2$,
and that the third polarization state with $\lambda_V = 0$ is now available.
For each plane-wave component with four-momentum $p_\mu$ inside the twisted state,
the orthogonality condition still holds: $p^\mu V_\mu(\lambda_V) = 0$.
For the transverse polarization states with $\lambda_V = \pm 1$, the expression \eqref{bbe-explicit} applies as it stands.
The longitudinal polarization now includes the time-like component and is described by the four-vector
\be
V_\mu (\lambda_V = 0) = \gamma (\beta, \bn_p)\,,
\ee
where $\bn_p = (\sin \theta_p \cos \varphi_p, \sin \theta_p \sin\varphi_p, \cos\theta_p)$ is the 3D unit vector along $\bp$
and $\gamma$ and $\beta$ are the standard relativistic kinematical quantities.
One can express the space-like part of this four-vector in the same basis ${\bm \chi}_\sigma$:
 \be
\bn_p =
 \sum_{\sigma=0,\pm 1} e^{-i\sigma \varphi_p}\,
 d^{1}_{\sigma 0}(\theta)  \,\bm \chi_{\sigma}\,.
\label{bbe-chi-0}
 \ee
In this way one can construct twisted Bessel states for a massive vector particle with arbitrary helicity $\lambda_V$
and angular momentum projection $m_V$.

\section{Collision of twisted photons}\label{section-collision}

\subsection{General features}

Let us first remind the reader of the general description of a scattering process
in which (some of) the initial or final particles are not plane waves.
Since we are not interested in calculation of numerical values of the cross sections,
we will describe the scattering amplitude omitting normalization factors.
We will consider the specific example of fusion of two initial particles into one final;
for a completely general treatment, see the review \cite{Kotkin:1992bj}.

In a scattering process, one is interested computing the scattering matrix elements from an initial to a final state.
In the plane wave case with the initial momenta $k_1$ and $k_2$
and the final momentum $p$ one has
\be
S_{PW}(k_1,k_2,p) \propto \delta^{(4)}(k_1 + k_2 - p)\cdot
{\cal M}(k_1,k_2;p) \,, \label{SPW}
\ee
where the invariant amplitude ${\cal M}$ is calculated according to the standard Feynman rules.
Then one calculates $|S_{PW}|^2$, regularizes the squares of delta-functions with finite volume and interaction time,
calculates transition probabilities, and integrates them over the final phase space to obtain the event rate.
Defining the flux of the colliding system, one extracts the cross section.

If the initial and final states are not plane waves but are described by appropriately normalized Fourier amplitudes
$\psi_1(\bk_1)$, $\psi_2(\bk_2)$, and $\psi_p(\bp)$, one calculates the scattering matrix element as
\be
S = \int d^3 k_1 d^3 k_2 d^3 p\, \psi_1(\bk_1)\psi_2(\bk_2)\psi_p^*(\bp) S_{PW}(k_1,k_2,p)\,.\label{S-tw}
\ee
We stress that the kinematical delta-function is present under the integral:
\be
S \propto \int d^3 k_1 d^3 k_2 d^3 p\, \psi_1(\bk_1)\psi_2(\bk_2)\psi_p^*(\bp)
\cdot \delta^{(4)}(k_1 + k_2 - p) \cdot {\cal M}(k_1,k_2;p)\,.\label{S-tw2}
\ee
One sees that integrations (partially) remove kinematical delta-functions,
rendering the resulting amplitudes and cross sections less singular.
Further analysis for collision of two twisted particles was performed in \cite{Ivanov:2011kk,Ivanov:2012na}.

\subsection{Producing spin-0 particle}

In this subsection we will calculate the production amplitude of a scalar particle $s$ with mass $M$
in collision of two twisted photons. We will first describe the final particle $s$ as twisted and deduce
the conservation law for the OAM projection, and then describe what happens if spherical harmonics
for the final particle are used. This calculation will prove that two twisted photons
have a non-zero overlap with the state with total angular momentum $J=1$.

We consider collision of two twisted Bessel photons $|E_1, \varkappa_1, m_1, \lambda_1\rangle$ and $|E_2, \varkappa_2, m_2, \lambda_2\rangle$ defined in the same frame and with respect to the same axis $z$.
The final scalar particle is described by the Bessel state $|E_p, K, m_s\rangle$ defined with respect to the same axis.
Let us choose the frame in which the longitudinal momenta of the colliding photons balance each other
as in Eq.~\eqref{balanced}: $k_{1z} = - k_{2z} \equiv k_z$ leading to $p_z =0$.
This is the closest one can get to the center of motion frame for twisted particles.
The energy conservation $E_1 + E_2 = E_p$ fixes the value of $K$ via
\be
\sqrt{\varkappa_1^2 + k_z^2} + \sqrt{\varkappa_2^2 + k_z^2} = \sqrt{K^2 + M^2}\,.
\ee
Since the longitudinal and transverse Fourier components factorize,
the scattering matrix element has now the following form:
\be
S \propto
\int d^2 k_{1\perp} d^2 k_{1\perp} d^2 p_{\perp}\,
a_{\varkappa_1 m_1}(\bk_{1\perp})\, a_{\varkappa_2, -m_2}(\bk_{2\perp})\, a^*_{K m_s}(\bp_{\perp})\cdot \delta^{(2)}(\bk_{1\perp} + \bk_{2\perp} - \bp_\perp) \cdot {\cal M}_s\,.
\label{S-gamgam-s}
\ee
Using the explicit expressions for the Fourier components in \eqref{S-gamgam-s},
we simplify the scattering matrix element further:
\be
S \propto \int d\varphi_1 d\varphi_2 d\varphi_p\, e^{i(m_1 \varphi_1 - m_2 \varphi_2 - m_s\varphi_p)}
 \delta^{(2)}(\bk_{1\perp} + \bk_{2\perp} - \bp_\perp) \cdot {\cal M}_s\,, \label{S-gamgam-s2}
\ee
where $|\bk_{1\perp}|=\varkappa_1$, $|\bk_{2\perp}|=\varkappa_2$, and $|\bp_{\perp}|=K$.

The invariant amplitude ${\cal M}_s$ is generated by the usual interaction Lagrangian
${\cal L}_{\gamma\gamma s} = g F^{(1)}_{\mu\nu} F^{(2) \mu\nu} s/4$
and has the following form
\be
{\cal M}_s = g\left[ (k_1k_2)(e_1e_2) - (k_1e_2)(k_2e_1)\right]\,.\label{Ms}
\ee
Contrary to the usual situation in the plane wave head-on collision,
for twisted photons in the Coulomb gauge, the polarization vectors \eqref{bbe-explicit} are not orthogonal
to the momenta of the counter-propagating photons, therefore we keep the second term in \eqref{Ms}.
This term can in fact be removed in a certain gauge, as we will discuss in the next subsection,
but we will not employ this transformation here.
The amplitude ${\cal M}_s$ can be evaluated for generic plane wave components of the two photons.
It is non-zero only for equal helicities $\lambda_1=\lambda_2=\lambda$ and has the form
(we use the shorthand notation $c_i \equiv \cos\theta_i$, $s_i \equiv \sin\theta_i$):
\be
{\cal M}_s = {g \over 2} E_1E_2 \delta_{\lambda_1, \lambda_2}\left[
e^{i(\varphi_1-\varphi_2)}(1-\lambda c_1)(1+\lambda c_2)
+ e^{-i(\varphi_1-\varphi_2)}(1+\lambda c_1)(1-\lambda c_2) - 2 s_1 s_2\right]\,.\label{Ms2}
\ee
Notice that it depends on the azimuthal angles
of the two photons only through their difference:
${\cal M}_s(\varphi_1, \varphi_2) = {\cal M}_s(\varphi_1- \varphi_2)$.
The paraxial case corresponds to $c_1 \to 1$, $c_2 \to -1$, and the expression simplifies
to $2 E_1E_2 \delta_{\lambda_1, \lambda_2} \exp[-i\lambda (\varphi_1-\varphi_2)]$

The constrained azimuthal integral \eqref{S-gamgam-s2} can be evaluated using the results
from Appendix A of \cite{Ivanov:2011kk}.
The integrations with respect to $\varphi_1$ and $\varphi_2$ eliminate the two-dimensional delta-function,
so that the integral is proportional to the sum
\be
\lr{{\cal M}_s} = \left[e^{i(m_1 \varphi_1 - m_2 \varphi_2 - m_s\varphi_p)} \cdot {\cal M}_s(\varphi_1- \varphi_2)\right]_{a}
+
\left[e^{i(m_1 \varphi_1 - m_2 \varphi_2 - m_s\varphi_p)} \cdot {\cal M}_s(\varphi_1- \varphi_2)\right]_{b}
\label{sumMs}
\ee
calculated at the following values of the azimuthal angles:
\bea
\mbox{configuration $a$:} && \varphi_1 = \varphi_p + \delta_1\,, \quad \varphi_2 = \varphi_p - \delta_2\,, \nonumber\\
\mbox{configuration $b$:} && \varphi_1 = \varphi_p - \delta_1\,, \quad \varphi_2 = \varphi_p + \delta_2\,,
\eea
where $\delta_1$ and $\delta_2$ are two inner angles of the triangle with the sides $\varkappa_1$, $\varkappa_2$, and $K$,
namely, the angles between $\varkappa_1$ and $K$ and between $\varkappa_2$ and $K$, respectively.
Substituting these values, one observes that $\varphi_p$ dependence disappears in ${\cal M}_s(\varphi_1- \varphi_2)$
and remains only in an overall factor $\exp[i(m_1 -m_2 -m_s)\varphi_p]$.
Integrating it over $\varphi_p$, we recover the conservation law of the
angular momentum projection:
\be
m_s = m_1 - m_2\,,
\ee
which holds for any helicity amplitude.
This conservation law is, of course, expected, since all three twisted states are defined with respect to the same axis $z$
and their interaction law conserves angular momentum.
Thus, by choosing appropriate $m_1$ and $m_2$,
one can achieve any value of $m_s$ including $m_s= \pm 1$.

We could have chosen to describe the final particle with spherical harmonics instead of Bessel states:
$\psi_p(\bp) \propto Y_{J m_s}(\theta_p, \varphi_p)$ in Eq.~\eqref{S-tw2}.
In the frame \eqref{balanced}, we would obtain the longitudinal momentum delta function
$\delta(p_z)$, which removes the integration over polar angles and sets $\theta_p = \pi/2$.
Since $Y_{1,\pm 1}(\theta_p = \pi/2, \varphi_p)$ is non-zero, we get a non-vanishing
amplitude of production of a scalar particle in the state $J=1$.

There is an elegant alternative way of reaching this conclusion by evaluating the coordinate space wave function
of the final scalar field produced in collision of two twisted photons.
One first repeats the above calculation in the plane wave basis for the final particle, that is, one evaluates
$S$-matrix element \eqref{S-gamgam-s} for a fixed final $\bp$:
\be
S(p) \propto \delta(E_1 + E_2 - E_p) \delta(k_{1z}+k_{2z}-p_k)\, e^{i(m_1-m_2)\varphi_p}\, \lr{{\cal M}_s}\,,
\ee
where $\lr{{\cal M}_s}$ defined in \eqref{sumMs} depends on $K = |\bp_\perp|$ but not on its azimuthal angle.
Next, one performs the Fourier transform to obtain the wave function of the produced scalar:
\be
\Psi(\br) = \int S(p)\, e^{i\bp \br}{d^3 p \over (2\pi)^3}\,.
\ee
The two delta-functions in $S(p)$ fix $p_z$ and $K$, and the remaining angular integration gives,
in cylindric coordinates,
\be
\Psi(\br) \propto e^{i(k_{1z}+k_{2z})z} e^{i(m_1-m_2)\varphi_r}
J_{m_1-m_2}(\rho K) \lr{{\cal M}_s}\,.
\ee
This wave function can be projected on any spherical harmonic, and in generic case it can
have a non-zero overlap with the states with $J=1$.

We conclude that there is no problem {\em per se} in constructing a pair of twisted photons
with a non-zero overlap with the $J=1$ state.
The only problem is that this state cannot be coupled to a spin-1 particle, which we will see in the following subsection.

\subsection{Producing spin-1 particle}

Production of vector particle $V$ in collision of two (twisted) photons
can be described by the same generic expression \eqref{S-tw2},
where the plane wave invariant amplitude ${\cal M}_V$ would describe the putative coupling
of two photons and a vector field.
This amplitude must originate from an interaction Lagrangian involving
$F_1^{\mu\nu}$ and $F_2^{\mu\nu}$ for the two photons,
as well as $V^*_\mu$ and derivatives.

If we require gauge and Lorentz invariance as well as Bose statistics for the two photons,
we will find that this plane wave amplitude is zero.
In fact, this statement is just the application of the Landau-Yang theorem to the plane wave case
in its formulation 1.
Therefore, the amplitude will remain zero even if weighted with any initial and final wave functions.

Since we mentioned in the introduction that some standard arguments can be not applicable to twisted photons,
we find it instructive to repeat this line of arguments paying special attention to the absence of the reference
frame in which the spin-1 particle is at rest.

The putative plane wave matrix element ${\cal M}_V(k_1,k_2;p)$ must be linear in $e_1^\mu$, $e_2^\mu$, and $V^*_\mu$.
It may contain several terms, and each of them, taken individually, does not have to be gauge invariant.
But remembering that all these terms originate from a gauge invariant Lagrangian,
we can take a convenient gauge choice and calculate each term in it.
We show below that, with a suitable gauge choice, each such term vanishes.

When writing down an interaction term in ${\cal M}(k_1,k_2;p)$,
we have at our disposal, apart from the polarization vectors,
the momenta\footnote{As a side remark, we notice that a twisted state of any particle is defined
in a specific frame and with respect to a specific axis.
It is tempting to conclude that we should also have at our disposal
an additional unit time-like four-vector $n_0^\mu$ such that $n_0^\mu = (1,0,0,0)$
would indicate the preferred frame. This expectation is erroneous.
Such a four-vector is needed to describe the twisted state, as a whole,
but not to specify the fundamental interactions of the plane-wave components of the twisted photon
and the spin-1 particle.}
$k_{1}^\mu$ and $k_{2}^\mu$, or equivalently their sum $p^\mu = k_{1}^\mu + k_{2}^\mu$
and difference $q^\mu = k_{1}^\mu - k_{2}^\mu$.
We can also contract vectors with the fully antisymmetric tensor $\epsilon_{\mu\nu\rho\sigma}$.
The Bose statistics of the photons requires that the expressions be symmetric under the simultaneous exchange
$k_1 \leftrightarrow k_2$ and $e_1 \leftrightarrow e_2$.
All these conditions imply that four possible structures
\be
(e_1 e_2)(V^* q)\,,
\quad
(e_1 k_2)(e_2 k_1)(V^* q)\,,
\quad
\epsilon_{\mu\nu\rho\sigma} e_1^\mu e_2^\nu V^{*\rho} p^\sigma\,,
\quad
\epsilon_{\mu\nu\rho\sigma} e_1^\mu e_2^\nu q^\rho p^\sigma \cdot (V^* q)
\label{wrong-structures}
\ee
are antisymmetric under the exchange of the two photons and cannot participate in the coupling.
One can only construct two structures which respect the Bose symmetry:
\be
T_1 = (e_1 p)(e_2 V^*) + (e_2 p)(e_1 V^*)\,,
\qquad
T_2 = \epsilon_{\mu\nu\rho\sigma} e_1^\mu e_2^\nu q^\rho V^{*\sigma} \,,
\label{new-structures}
\ee
In the usual case of the plane wave collision, in the rest frame of the massive spin-1 particle
with the photons colliding along axis $z$,
one can use gauge freedom to make $e_1^\mu$ and $e_2^\mu$ purely transverse 2D vectors.
As a result, both structures $T_1$ and $T_2$ vanish:
the former vanishes due to $(e_{1,2}p) = 0$, and the latter disappears because
all four vectors involved in the product have zero time-like components.

Let us stress once again the very important fact that the above arguments are valid always,
irrespective of the non-trivial spatial distribution of the colliding photons.
Indeed, the coordinate wave functions are just the envelopes with which one integrates the core quantity,
the plane-wave scattering amplitude.
The structures \eqref{wrong-structures} were found antisymmetric and were eliminated
on the basis of their non-compliance with the Bose symmetry of the plane-wave photon-annihilating {\em operators},
not due to the coordinate wave function properties.

At the first look, this simplifying trick does not apply for the case of twisted photons.
Each of the two photons is a superposition of many plane wave components,
and the structures $T_1$ and $T_2$ are calculated for each pair of $k_1$ and $k_2$.
Even if one performs a gauge transformation of the form \eqref{gauge-tr} to make
$e_1'$ orthogonal to one specific $k_2$, this orthogonality will certainly fail for all other plane wave components in the second twisted photon.
Thus, it seems that we cannot immediately set $T_1$ to zero in the integrand of \eqref{S-tw2}.
As for the structure $T_2$, although it is possible to make $q = k_1 - k_2$ purely spatial
by performing a longitudinal boost and setting $E_1 = E_2$,
the longitudinal momenta will not be balanced in this frame, $k_{1z}+ k_{2z} = p_z \not = 0$,
and as a result $V_\mu(\lambda_V = 0)$ will contain a time-like component.
Therefore, it seems that one cannot just disregard this structure.

Nevertheless, the same arguments can be employed here as well, although in a slightly
different manner.
We use gauge freedom to set structures $T_1$ and $T_2$ to zero not for each plane wave pair
of the colliding photons, but for each plane wave component of the final particle wave packet.
Through the momentum delta-function, it fixes the plane wave component of the second
photon for each plane wave component of the first one.
Namely, let us pick up the first term in the structure $T_1$ and use the integration over $\bk_2$
to remove the momentum delta-function:
\be
S \propto \int d^3 p\, \psi_p^*(\bp)\int d^3 k_1 \, \psi_1(\bk_1)\psi_2(\bp-\bk_1)
\cdot (e_1 p) (e_2V^*) \cdot f(k_1, p-k_1; p)\,.\label{S-tw3}
\ee
For a given final particle momentum $p$, one can always apply the gauge transformation
\eqref{gauge-tr} to set $(e_1 p) = 0$ for each $\bk_1$.
This gauge-fixing condition, which depends on $k_1$, may seem unusual,
but it originates, in the coordinate space, from the familiar shift
$A_\mu(x) \to A_\mu(x) + \partial_\mu \alpha(x)$ applied to the twisted photon.
In a similar way, the second term of the structure $T_1$ is removed by performing
the gauge transformation on the second photon to set $(e_2 p) = 0$.
The result is that, within this gauge fixing choice, the structure $T_1$ vanishes
for each final momentum $p$ individually.

It is remarkable that, with the same gauge fixing choice,
the structure $T_2$ vanishes as well. Indeed, since $e_1$ and $e_2$ are orthogonal to both $p$ and $q$,
the vector
\be
\epsilon_{\mu\nu\rho\sigma} e_1^\mu e_2^\nu q^\rho \propto p_\sigma\,.
\ee
But the structure $T_2$ represents the scalar product of this vector with $V^*$.
Since $(p V^*) = 0$, this structure disappears.

We conclude that, just as in the plane wave case, there is no structure which could couple
a pair of twisted photons with a spin-1 particle in any state.

\section{Conclusions}

There exist two formulations of the Landau-Yang theorem:
one forbidding coupling of two real photons to a spin-1 field (formulation 1),
and the other forbidding two photons to be in a state with total angular momentum $J=1$ (formulation 2).
They are often formulated and proved for two photons with zero total momentum,
and in particular for back-to-back plane wave photons.

But what does the Landau-Yang theorem say about the system of two {\em twisted} photons,
which carry orbital angular momentum
with respect to their propagation direction? Can one extend it, modifying its formulation?
What is actually forbidden by the Landau-Yang theorem in this case?

Answering these simple questions may lead to some confusion,
especially taking into account that standard proofs begin by switching
to the center of motion frame, which is not well defined for twisted particles.
In this paper, we discussed this question, accurately disentangling several parts of the problem,
and clarified the confusing aspects.
We showed that two twisted photons {\em can} have non-zero overlap with the total $J=1$ state.
In particular, collision of such photons can produce a massive scalar particle in this angular momentum state.
This proves that formulation 2, which is initially given only for plane-wave photons,
{\em cannot} be extended to twisted photons or to any other photon states with non-trivial
spatial wave functions, even if one
selects the reference frame where the average momenta of the two photons balance each other.

What is actually forbidden --- and twisted photons are no exception here ---
is production of spin-1 particles in collisions of two photons described by arbitrary wave packets.
Two photons can be in the state $J=1$, and so can a spin-1 particle ---
but the two systems just cannot be coupled in a way which respects Lorentz and gauge invariance and Bose statistics.
We repeated the standard arguments adapting them to the case where no center of motion frame exists.
The net result is that no miracle happens.

It is intriguing to check whether collisions of twisted particles,
with their extra degrees of freedom absent in the plane wave collisions, can lead to novel phenomena
or provide additional insights in situations where the Landau-Yang theorem is not applicable.
These cases include two-photon processes when one or both photons are virtual
or collisions of two distinct vector particles, such as exclusive photoproduction
of the vector meson $f_1(1285)$ \cite{Kochelev:2009xz,Dickson:2016gwc} or its decay to $\rho\gamma$
\cite{Osipov:2017ray}.
Landau-Yang theorem is also avoided for pairs of real gluons \cite{Beenakker:2015mra,Cacciari:2015ela}.
Since two gluons can be in color antisymmetric state, the structures \eqref{wrong-structures}
are now allowed. Nevertheless, it turns out that $gg\to q\bar q$ production in
the color-octet ortho-quarkonium state is still zero at the Born level,
so that the first non-trivial contribution appears at the next-to-leading order of QCD \cite{Beenakker:2015mra}.
It would be interesting to see if the result changes for twisted gluons.

\section*{Acknowledgements}
We are grateful to Ilya Ginzburg
for critical reading and valuable comments.
I.P.I. wishes to thank the Institute of Modern Physics, Lanzhou, China, for financial support and hospitality during his stay.
I.P.I. acknowledges funding from the Portuguese
\textit{Fun\-da\-\c{c}\~{a}o para a Ci\^{e}ncia e a Tecnologia} (FCT) through the FCT Investigator
contract IF/00989/2014/CP1214/CT0004
and through the projects UID/FIS/00777/2013, UID/FIS/00777/2019,
CERN/FIS-PAR/0004/2017, and PTDC/FIS-PAR/29436/2017,
which are partially funded through POCTI (FEDER), COMPETE, QREN, and the EU.
I.P.I. also acknowledges the support from National Science Center, Poland,
via the project Harmonia (UMO-2015/18/M/ST2/00518).
V.G.S. is supported by the Russian Science Foundation (Project No. 17-72-20013).
P.M.Z. is supported by the National Natural Science Foundation of China (Grant No. 11975320)

\end{document}